\documentclass[9pt,twocolumn,twoside]{osajnl}

\journal{ol} 

\setboolean{shortarticle}{true}

\title{In-situ probing and stabilizing the power ratio of electro-optic-modulated laser pairs based on VIPA etalon for quantum sensing
}

\author[1,2,3,$\dagger$]{Guochao Wang}
\author[1,2,$\dagger$]{Mingyue Yang}
\author[1,2]{Enlong Wang}
\author[1,2]{Xu Zhang}
\author[1,2]{Aiai Jia}
\author[1,2]{Lingxiao Zhu}
\author[1,2]{Shuhua Yan}
\author[1,2,*]{Jun Yang}

\affil[1]{National University of Defense Technology, College of Intelligence Science and Technology, Changsha, Hunan 410073, China}
\affil[2]{National University of Defense Technology, Interdisciplinary Center for Quantum Information, Changsha, Hunan 410073, China}
\affil[3]{High-tech Institution of Xi’an, Xi’an 710025, China}
\affil[*]{Corresponding author: jyang@nudt.edu.cn, zhulingxiao31@163.com}
\affil[$\dagger$]{These authors contributed equally to this work.}

\begin{abstract}
Monitoring and stabilizing the power ratio of laser pairs is significant to high-precision atom interferometers, especially as the compact electro-optic modulated all-fiber laser system prevails. 
In this Letter, we demonstrate a novel method to in-situ probe the relative power of laser pairs and to stabilize the power ratio of two Raman lasers using a high-dispersion virtually imaged phased array (VIPA) etalon.
Sub-microsecond resolution on probing laser power transformation during atom interferometer sequence is achieved and the power ratio of two Raman lasers (PRTR) is tightly locked with high bandwidth despite of environmental disturbances, showing an Allan deviation of $4.39\times 10^{-5}$ at 1000~s averaging time.
This method provides a novel way to stabilize the PRTR and diagnose the multi-frequency laser systems for atom interferometers and could find potential application in broad quantum sensing scenarios.
\end{abstract}

\begin{document}

\maketitle

Atom interferometers (AIs) have played an increasingly important role in state-of-the-art fundamental physics tests and quantum sensing scenarios with applications in a broad range from the test of Einstein’s equivalence principle~\cite{rosi2017quantum}, determination of the Newtonian gravitational constant~\cite{rosi2014precision} to the precise measurement of gravity~\cite{peters2001high}, gravity gradient~\cite{sorrentino2012simultaneous} and rotations~\cite{canuel2006six}. For a typical Raman-type AI, a compact and effective method for generating a pair of highly coherent Raman lasers is through electro-optic modulation, which exhibits low phase noise but suffers from the drift in modulation depth due to environmental disturbances~\cite{luo2019compact, dotsenko2004application}. This drift leads to instability of the power ratio of two Raman-lasers (PRTR), which will induce unwanted AC Stark shift~\cite{zhu2017measurement} in AIs and severely undermine the measuring precision in the case of short interrogation time or long-term operation. The probing and stabilization of this PRTR is therefore of practical importance to a high-precision and transportable AI.

Various approaches have been proposed and demonstrated for the probing and stabilization of PRTR, including the use of a Fabry–Pérot (FP) cavity~\cite{luo2019compact}, a frequency spectrum analyzer (FSA)~\cite{zhu2017measurement} and Schottky diode detectors (SDD)~\cite{wang2020new}. The FP method extracts the PRTR by scanning the cavity length for sequential power detection, while the FSA and SSD methods rely on beating the Raman beams with an additional reference beam and the induced beatnotes are used for servo locking the PRTR. 
However, the FP method fails to provide real-time monitoring and the FSA method is limited by the acquisition rate and maximum 10~Hz bandwidth for PRTR control. Finally, the SDD method provides several tens of kHz bandwidth at the cost of complicated optical and electronic processing. 
The latency or the complexity of the above mentioned methods prevent the applications of electro-optic modulated Raman lasers in AIs for field applications.
An alternative and intuitive solution is to spatially separate the laser pair and directly monitor and fast control their power ratio. However, considering the frequency gap to be around 6.8~GHz in Rb, it is difficult for traditional dispersion devices such as gratings and prisms to spatially separate Raman laser pair with large enough distance for probing. Following this intuition, we focused on a type of spectral dispersion etalon named the virtually imaged phased array (VIPA), which can separate the spectrum in sub-GHz regime~\cite{shirasaki1996large, xiao2004dispersion}. 
Due to its simple structure, polarization insensitivity and high dispersion, VIPA has been initially used in wavelength division multiplexing~\cite{xiao2005eight}, optical communication~\cite{xiao20042}, pulse shaping~\cite{supradeepa2008femtosecond}, tomography~\cite{zhang2017three}, etc.

In this Letter, we demonstrate an in-situ, high-precision and low-drift method to probe and stabilize the PRTR using a VIPA etalon. 
Thanks to the high-dispersion feature of the VIPA,  we have spatially separated the electro-optic modulated light (wavelength 780~nm, frequency gap 6.8~GHz) by 3.5~mm and realized in-situ monitoring and control of the power ratio between laser pairs for AI. 
As direct optical detection is used, this method has no latency in probing and circumvents the additional reference beams and complicated circuity.
Sub-microsecond resolution on probing, robust locking and reinforced long-term stability of the PRTR is realized.
This method can find direct application in EOM-modulated quantum sensing scenarios where the power ratio between laser pairs is an impacting factor.

\begin{figure}[t!]
   \centering
   \includegraphics[width=1\linewidth]{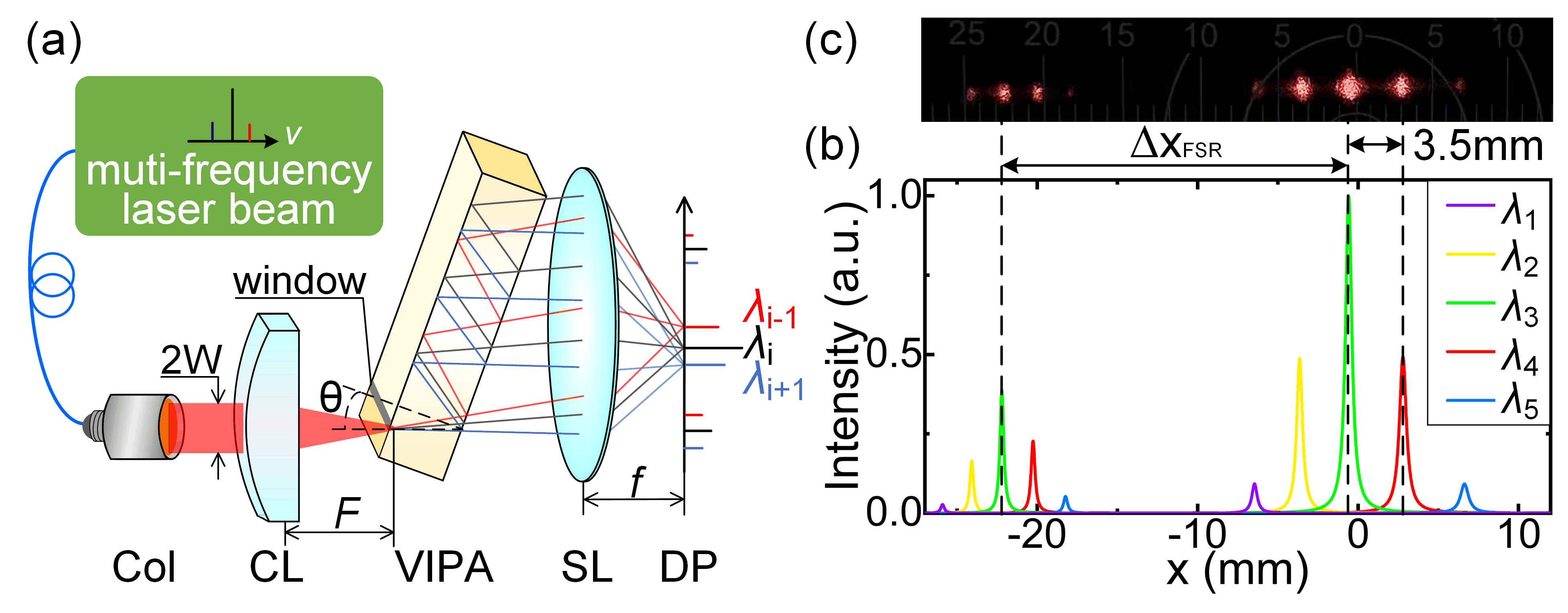}
   \caption{(a) Schematic of the VIPA-based dispersion. Col: collimator, CL:cylindrical lens, SL: sperical lens, DP: detection plane. (b) Captured photograph of VIPA-dispersed laser beams on the DP. (c) Simulation results under realistic parameters: $W = 0.6$~mm, $F = 62$~mm, $f = 2000$~mm and $\theta = 1$\textdegree{}.}
   \label{fig1}
\end{figure}

The schematic of the VIPA-based dispersion is shown in Fig.~\ref{fig1}~(a). 
The core element is the VIPA etalon which consists of two parallel facets. 
The front facet is coated with nearly 100\% reflective film except for a narrow window that allows the incidence of the laser beam with a small angle (normally < 5°), while the back facet is coated with a partially reflectivity film (> 95\%). 
A vertically compressed input beam illuminates onto the narrow window through a cylindrical lens and undergoes multiple reflections inside the VIPA.
The laser components with the same frequency are displaced in parallel and exit from the back facet at each reflection point, while those with different frequencies deviate by an angle due to different refractive index. 
Finally, the dispersed laser beams are focused by a spherical lens and converge to different spots on the focal plane corresponding to their respective frequency components.
To be noted, similar to an FP etalon, the free spectral range (FSR) of the VIPA is characterized as a certain spacing gap.

According to Ref.~\cite{xiao2004dispersion}, the intensity of the output light $I_{\mathrm{out}}$ on the focal plane through the VIPA-based dispersion system can be expressed as:
\begin{equation}
\begin{split}
I_{\mathrm{out}}(x_{F},\lambda) &\propto |E_{\mathrm{out}}(x_{F}, \lambda)|^{2} \\
&\propto \exp\left(-\frac{2f^{2}x_{F}^{2}}{F^{2}W^{2}} \right) \frac{1}{(1-Rr)^{2}+4Rr\sin^{2}(k\Delta/2)},
\end{split}
\end{equation}
where $E_{\mathrm{out}}$ is the electric field intensity, $x_{F}$ the position along the beam separation direction on the focal plane, $\lambda$ the wavelength, $k=2\pi/\lambda$ the wave vector and $W$ the radius of the collimated beam. $R$ and $r$ are the reflectivity of the front and back facets, respectively; $F$ and $f$ are the focal length of the cylindrical and the spherical lenses, respectively. $\Delta$ is the optical path difference:
\begin{equation}
\Delta = 2t\cos(\theta)- \frac{2t\sin(\theta)x_{F}}{F} - \frac{t\cos(\theta)x_{F}^{2}}{F^{2}} ,
\end{equation}
where $t$ is the thickness of the VIPA and $\theta$ is the incident angle.
Assuming a fixed input beam, the output spot interval can be affected by $\theta$, $f$, $F$, $W$ and $t$.

Figure~\ref{fig1} (b) and (c) show a captured photograph on the focal plane and the output light field distributions based on the simulations, respectively.
In order to demonstrate the target scenario of Raman laser pairs for Rb AI, 
the multi-frequency laser beam is generated by phase-modulating a single frequency laser at wavelength $\lambda = 780.24$~nm with an electro-optic modulator (EOM). The modulation frequency is 6.834~GHz and the modulation depth is 17~dBm, resulting in observable 1st and 2nd order sidebands along with the carrier.
As the on-shelf VIPA (model: OP-6721-1686-4) is applied in our demonstration, the thickness and the FSR are manufactured as $1.686$~mm and $60$~GHz, respectively.
Besides the parameters listed in the caption of Fig.~\ref{fig1}, for the simulation, the intensity ratio of the carrier to the 1st and 2nd order sidebands is set as  $1:0.5:0.1$. 
Under those parameters, we simulate the spatial separation  and the relative intensity ratio of the dispersed light beams with different frequency components and the results are shown in  Fig.~\ref{fig1}~(c). 
The simulated light spot interval of 3.5~mm agrees with the real-life separation photographed on a scaled paper putting at the focal plane of the spherical lens (Fig.~\ref{fig1}~(b)). 
Besides, the dimer spots on the left side of Fig.~\ref{fig1}~(b) are formed by another spectral order of the VIPA dispersion and the distance from the main spot corresponds to the FSR in the spatial domain.  
In addition, the intensity ratio of the separated beams recorded by a power meter is in agreement with the simulation results, confirming the effectiveness of the VIPA dispersion for multi-frequency laser beam.

\begin{figure}[t!]
   \centering
   \includegraphics[width=1\linewidth]{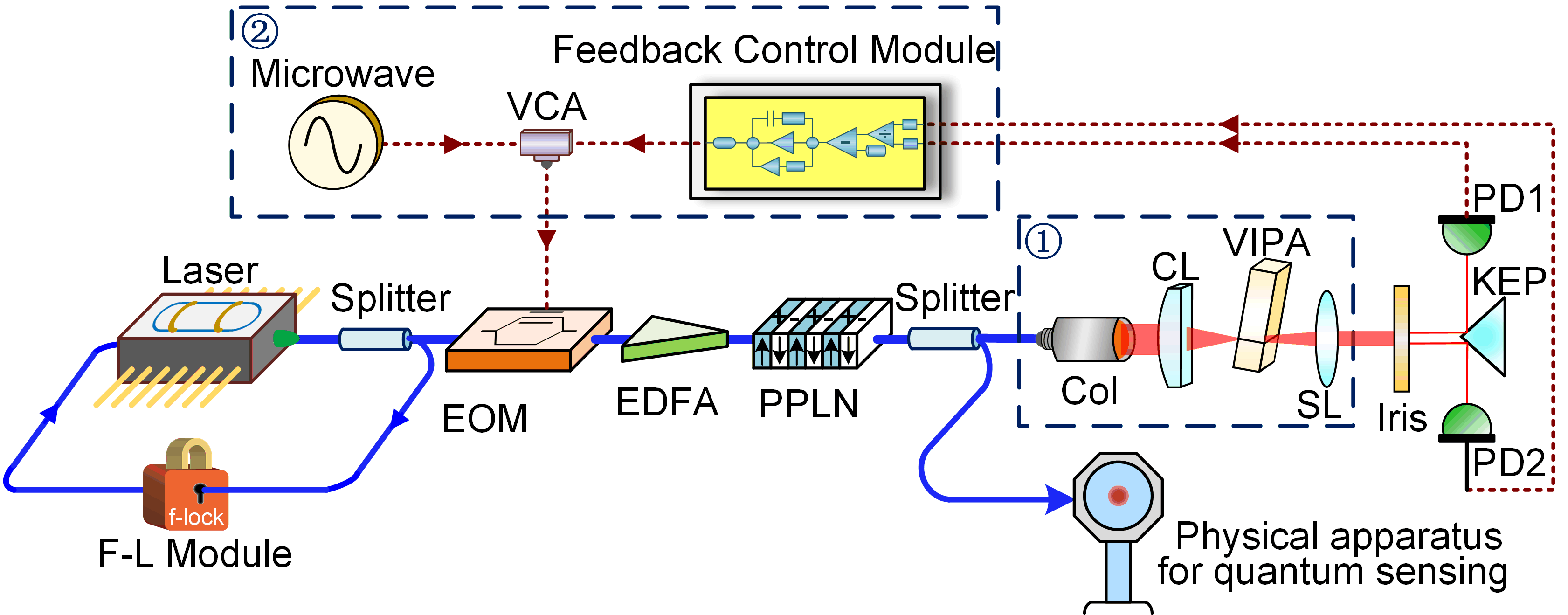}
   \caption{Schematic diagram of the PRTR probing and stabilizing system. F-L Module: frequency-locked module, VCA: voltage-controlled attenuator, EDFA: erbium doped fiber amplifier, PPLN: periodically poled lithium niobate, PD: photodetector, KEP: knife edge prism.}
   \label{fig2}
\end{figure}

Figure~\ref{fig2} shows the experimental setup of a quantum sening scenario where the VIPA method is used for probing and stabilizing the PRTR. 
A narrow-linewidth fiber laser at 1560~nm is frequency controlled by a frequency-locked module, where flexible frequency shift is demanded for AI as well. The partial of the laser source is then phase modulated, power amplified and frequency-doubled to a multi-frequency laser source at 780~nm with an EOM modulator, an EDFA and a PPLN module, respectively. Among these multi-frequency lasers, the two main components as a laser pair, generated by second harmonic genenration and sum frequency generation from the carrier and the +1st sideband of the 1560~nm EOM and termed as center frequency and +1st order frequency, manage to cover all different laser pairs for atom interferometer sequentially in time division, such as the cooling laser and the repumping laser for magneto optical trap (MOT) and polarization gradient cooling (PGC), Raman laser pair for Raman process, etc. The majority of the laser power (99\%) is directed to the physical apparatus for quantum sensing, while the remaining part (1\%) is sent to the VIPA module. On the basis of VIPA dispersion, the laser pair, targeted at the center frequency and +1st order frequency with 6.4-6.9 GHz frequency interval, is spatially separated into two beams.
In order to facilitate the detection of this laser pair, an iris diaphragm and a knife edge prism (Thorlabs MRAK25-P01) are dedicated to screen the laser pair out and make the respective beam far apart. Two parrallel photodetectors (Thorlabs PDA36A-EC) are used to detect the respective intensity of the two beams. The detected signals are further processed in the control module for probing the power ratio of the center frequency and +1st order frequency (PRCF) and locking the PRTR. This locking loop is accomplished through controlling the input RF power on the EOM by a VCA for the modulation depth tuning. Once the locking is on, the PRTR is stabilized to the set value.

\begin{figure}[t!]
   \centering
   \includegraphics[width=0.8\linewidth]{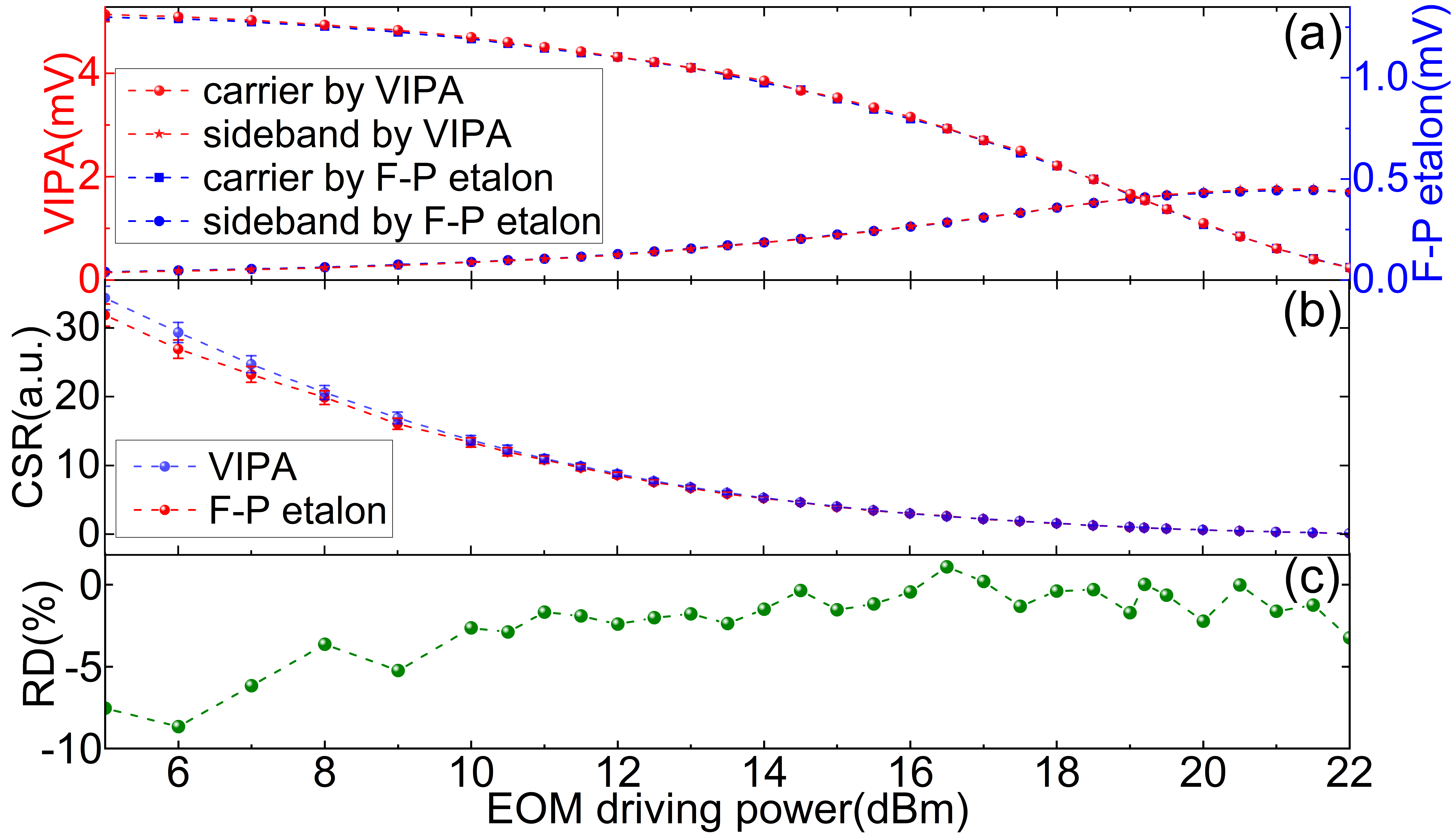}
   \caption{Comparison of the CSR between the VIPA dispersion method and the FP method with different driving powers imposed on EOM. (a) The carrier power and sideband power measured with two methods. (b) Calculated CSR and the error bar. (c) CSR relative difference between two methods.}
   \label{fig3}
\end{figure}

To verify the accuracy of probing the power ratio with our method, we directly used the electro-optic modulated lasers at 780.24 nm as in Fig.~\ref{fig1} and compared the VIPA-based system with a traditional FP method using a commercial scanning FP etalon (Thorlabs, SA200-5B).  
Figure~\ref{fig3} (a) shows the detected relative powers of the carrier and +1st sideband in two different cases as EOM modulation depth increases, where all the voltage data are captured on a digital oscilloscope (Tektronix, MSO64B). The left and right axes represent the values of the VIPA method and the FP method, respectively. To guide the view, the two axes are scaled by a constant, resulting in a perfect coincidence on plotted results. The power ratio of the carrier and +1st sideband (CSR) with the two methods are plotted in Fig.~\ref{fig3} (b), which shows good agreement within error bars. That ratio decreases as the EOM driving power (modulation depth) increases.
To further prove the consistency of these two methods, we define the relative difference ($RD$) in the CSR as $RD = 2~ (\mathrm{CSR}_{\mathrm{VIPA}}-\mathrm{CSR}_{\mathrm{FP}})/(\mathrm{CSR}_{\mathrm{VIPA}}+\mathrm{CSR}_{\mathrm{FP}})$.
The $RD$ in percentage is shown in Fig.~\ref{fig3} (c), which is within $\pm 3\%$ when the EOM driving power is in the range of $10\sim 21$~dBm. The $RD$ is relatively higher when the driving power is small as a result of larger measurement error in the case of a weaker sideband signal. This comparison reveals that the VIPA-based system behaves as accurate as the FP etalon. 
However, with VIPA we can perform continuous and real-time probing instead of scanning measurement, which is an obvious advantage over the FP method.

\begin{figure}[t!]
   \centering
   \includegraphics[width=0.8\linewidth]{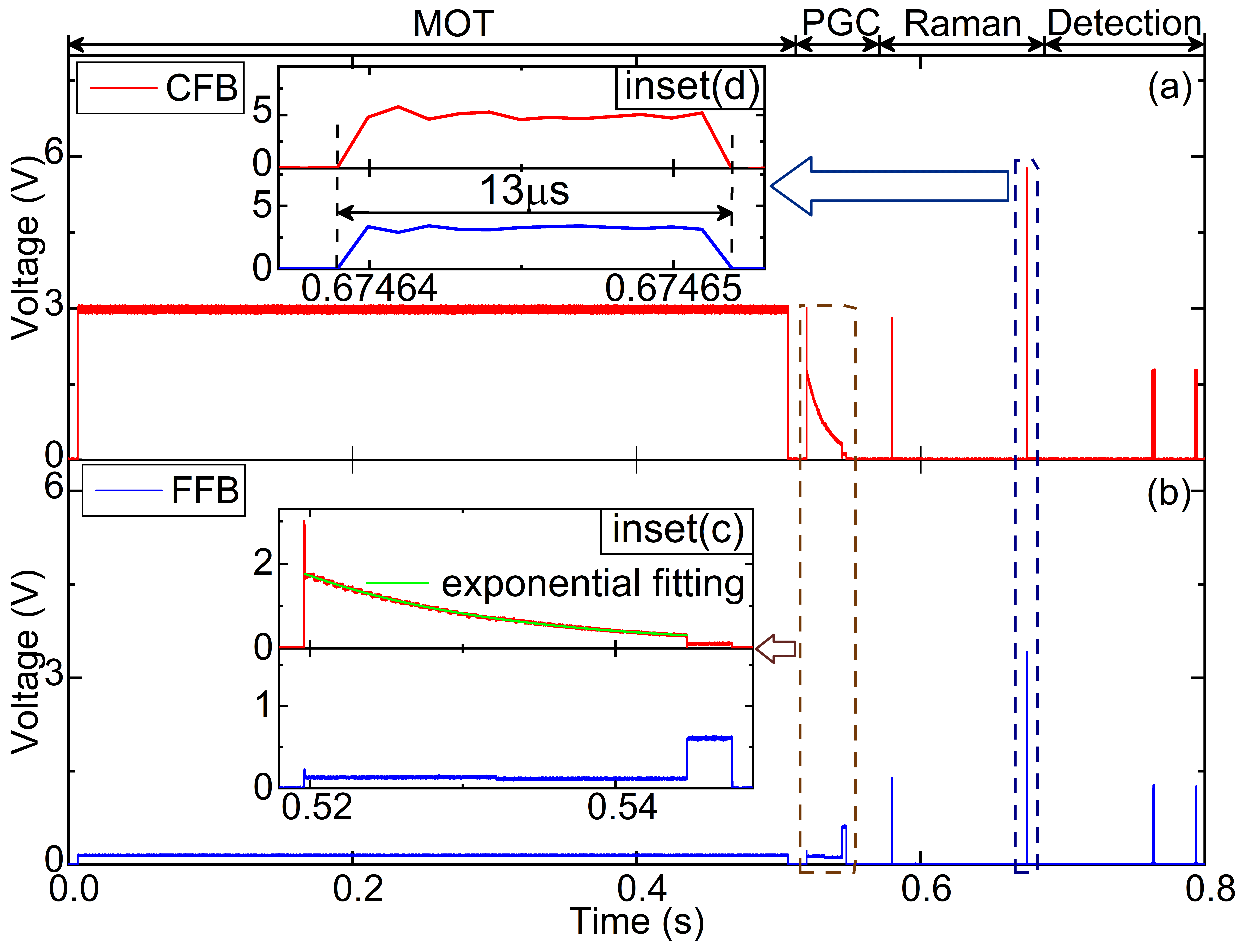}
   \caption{In-situ probing of the PRTR in an AI sequence with the VIPA-based system. (a) The detected power of CFB in red. (b) The detected power of FFB in blue. The inset (c) and the inset (d) are enlarged detailed figures of the PGC and Raman stages.}
   \label{fig4}
\end{figure}

In light of the in-situ and continuous probing, we tried to use the VIPA-based system to monitor the power ratio of laser pairs in the experimental sequence of an AI, where the laser source used is the one described in Fig.~\ref{fig2} and the concerned power ratio is exactly the PRCF. The in-situ monitoring on the laser pairs during one cycle of AI is shown in Fig.~\ref{fig4}, which exhibits a duration of about 0.8 second and contains four stages: the MOT, PGC, Raman and detection. The relative powers of two seperated laser beams from the laser pair, denoted as the center frequency beam (CFB) and the +1st-order frequency beam (FFB), are measured by the mentioned oscilloscope with a sampling rate of 1~M/s (limited by the sampling storage), and the results are given in Fig.~\ref{fig4} (a) and (b), respectively.
Among all four stages, the PGC and the Raman are of particular interest to the diagnose of the atom interferometer sequence, and are detailed in the insets (c) and (d) of Fig.~\ref{fig4}, respectively.
In the PGC stage, the cooling beam (CFB in red) intensity exponentially decay while the repumping beam (FFB in blue) intensity keeps unchanged. The fitting curve of the detected cooling beam power shows full agreement with the designed exponential decay function. In the end of the PGC stage, the repumping beam intensity rise up while that of the cooling beam drops to zero, which corresponds to pumping all the cold atoms to the target state. 
Next, during the Raman stage there are two Raman $\pi$-pulses, the first of which is used for velocity selection and the second is to measure the Rabi oscillation. The second pulse draws special attention since it has the shortest pulse width in the whole sequence and its accuracy affects the determination of the Rabi frequency. Inset (d) of Fig.~\ref{fig4} shows the zoom of the pulse shape, which has a width of 13~$\mu$s and the average power ratio of the CFB (corresponding to Raman laser 1) and the FFB (corresponding to Raman laser 2) is about 1.51, almost the same as the set value.
Those results confirm that the VIPA-based method is estimated to resolve the experimental sequence of AI in a timescale of sub-microsecond.

\begin{figure}[t!]
   \centering
   \includegraphics[width=0.8\linewidth]{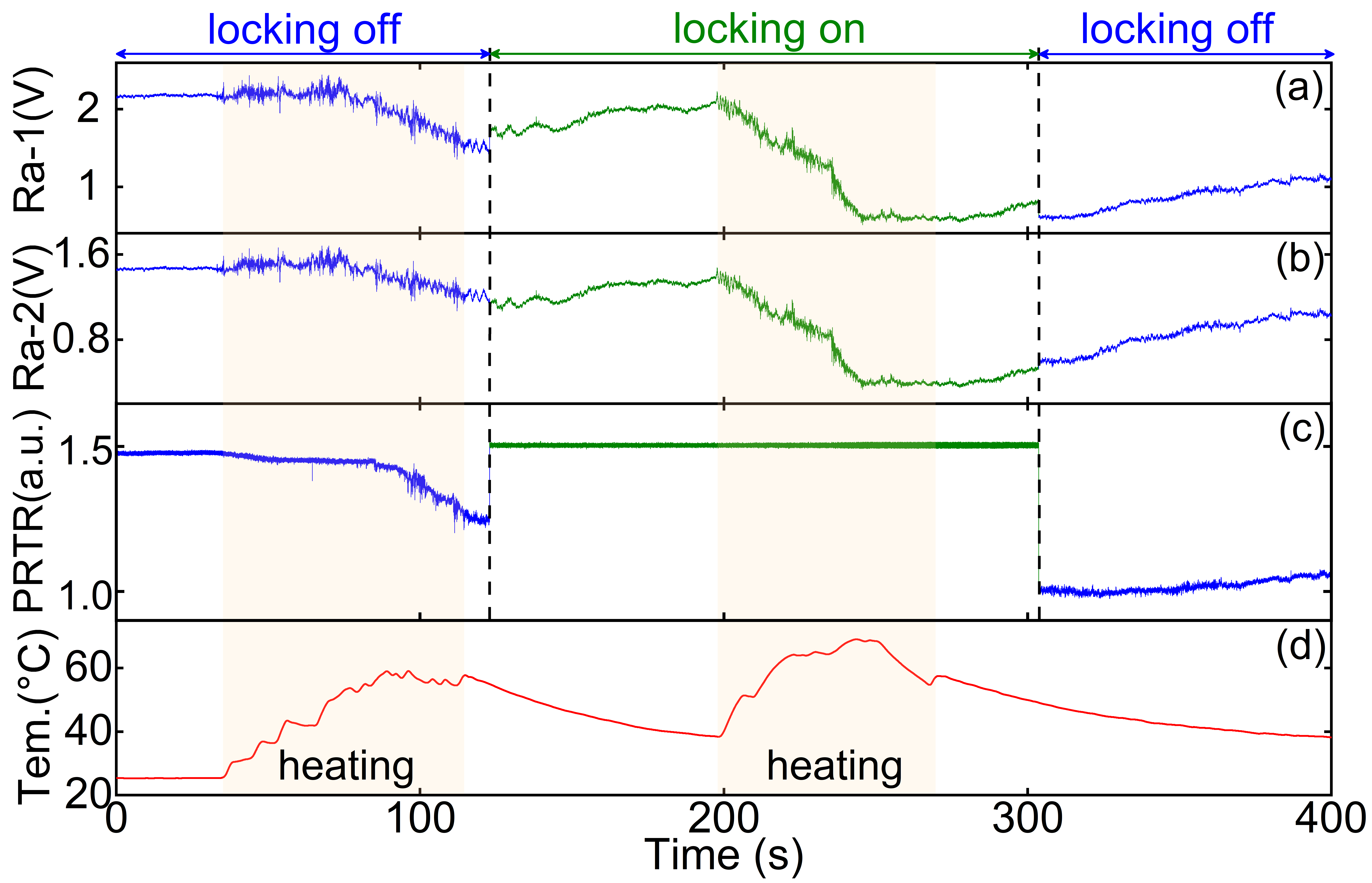}
   \caption{PRTR results in the open-loop and closed-loop regime under temperature disturbance. The detected power of Raman laser 1 (Ra-1) (a) and   Raman laser 2 (Ra-2) (b) and their power ratio (c) as temperature changes (d). The green and blue lines correspond to two switching states (PRTR locking on and off) and the shaded area indicates the heating-up.}
   \label{fig5}
\end{figure}

Making use of the favorable accuracy and temporal resolution on the power ratio with the VIPA-based method, we further explored the capability of the PRTR locking and its robustness against environmental disturbance with the help of a home-made feedback control module. To prove the anti-disturbance ability of our locking system, the EOM attached with a temperature sensor probe (Thorlabs-TSP01) was enclosed in a sealing cover, hot air flow from a hair-dryer was blew into the sealing cover to heat up the EOM. Basically, the whole test process was divided into three segments according to locking on and off, during which two rounds of heating up happened. 
Figure~\ref{fig5} (a)\textasciitilde(d) give the 400~s-duration results of the relative power of Raman laser 1 and Raman laser 2, PRTR and temperature, respectively. In Fig.~\ref{fig5} (c), when the locking loop is off, the PRTR initially varies slowly and slightly, but gradually drops down from 1.48 to 1.25 as the heating-up is on. It is notable that the PRTR change exhibits a $5\simeq50$ seconds delay versus temperature variation due to the warm-up time of the sealing cover. Once the locking is activated, the PRTR jumps up to the set value at 1.5 immediately, and is stabilized constantly with standard deviation of $2.96 \times 10^{-3}$, even if there is a temperature variation of 35~\textcelsius. At the last segment, the PRTR falls close down to 1 and rises slowly with the temperature recovering to the room temperature. 
These results testify the robustness of the PRTR locking against environmental disturbances.

\begin{figure}[t!]
   \centering
   \includegraphics[width=0.95\linewidth]{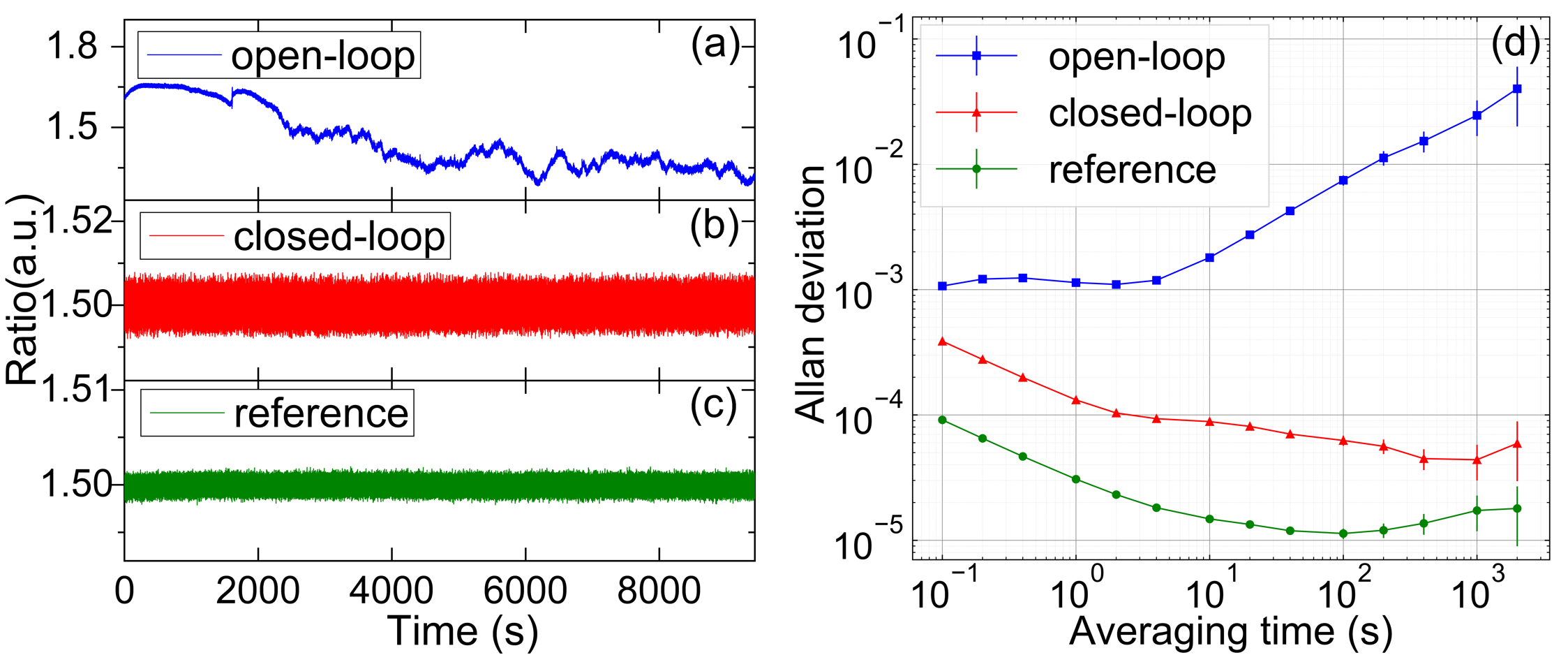}
   \caption{The long-term stability of the experiment system. (a) The PRTR of the open-loop regime. (B) The PRTR of the closed-loop regime, (c) The reference signal for PRTR locking. (b) Allan deviations of the recorded data.}
   \label{fig6}
\end{figure}

Finally, we evaluated the long-term stability of the PRTR locking. We recorded the PRTR of the open loop and closed loop for 10,000~s at a sampling rate of 0.01 s as well as the reference signal. The recorded results are shown in Fig.~\ref{fig6} (a), (b) and (c), respectively.   
During the whole measurement, the closed-loop PRTR and the reference signal have a standard deviation of $1.86\times 10^{-3}$ and $4.34\times 10^{-4}$, while that of the open-loop is $1.13\times 10^{-1}$, two orders of magnitude higher than the closed-loop.  
The Allan deviations of the above recorded data are shown in Fig.~\ref{fig6} (d). In the closed loop regime, the Allan deviation is improved  from $1.14\times 10^{-3}$ to $1.32\times 10^{-4}$ at 1 s averaging time, one order of magnitude better than that of the open loop regime increasing as a result of the drift over time. It is worth noting that the lowest Allan deviation of the closed loop reaches $4.39\times 10^{-5}$ at 1000 s averaging time, nearly three orders of magnitude lower than the open loop regime. As a benchmark, the reference signal reaches its lowest Allan deviation of $1.13\times 10^{-5}$ at 100 s averaging time, which ensures long-term stability of the locking system.
Compared to Ref.~\cite{zhu2017measurement}, our result with respect to the Allan deviation of PRTR is one order of magnitude better. Supposing an interrogation time of $50$~ms for a compact and transportable AI, we estimate that the error introduced by the PRTR in the g-measurement would be reduced from $10^{-7}$~g to $10^{-10}$~g by calculation. This improvement may suppress the effect of the PRTR instability to a level negligible to a compact and high-precision AI.

In conclusion, we have demonstrated an in-situ monitoring and accurate stabilizing method of the relative power of laser pairs in AI with a VIPA-based dispersion system.
Those coupled laser pairs can be spatially separated by more than 3~mm. 
Importantly, this system can respond to the control sequence of laser power transformation in sub-microsecond timescale, which is rarely achieved with other methods. 
Moreover, thanks to high temporal resolution and systematic robustness, the system can steady lock the PRTR at the presence of temperature disturbance up to 35~\textcelsius, and the long-term  locking stability for PRTR has been improved by almost three orders of magnitude in terms of Allan deviation.
The proposed method and system provide not only a novel solution for fast PRTR probing and robust stabilization, but also a practical way to diagnose the EOM-modulated laser system and debug the working sequence of atom interferometer, which could find direct application in quantum sensing scenarios for field applications where EOM-based generation of laser pairs is involved.
\paragraph{Acknowledgment} This work was supported by National Natural Science Foundation of China (Grant No. 12004428), Natural Science Foundation for outstanding young of Hunan Provincial, China (Grant No. 2021JJ20047), China Postdoctoral Science Foundation (Grant No. 2020M683729), and Science Foundation of Hunan Provincial, China (Grant No. 2021JJ30774). We acknowledge Dr. Liang Hu for a critical reading of this manuscript.
\paragraph{Disclosure} The authors declare no conflicts of interest related to this paper.

\bibliography{OL_VIPA_submit}



\ifthenelse{\equal{\journalref}{aop}}{%
\section*{Author Biographies}
\begingroup
\setlength\intextsep{0pt}
\begin{minipage}[t][6.3cm][t]{1.0\textwidth} 
  \begin{wrapfigure}{L}{0.25\textwidth}
    \includegraphics[width=0.25\textwidth]{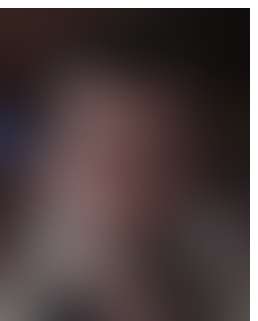}
  \end{wrapfigure}
  \noindent
  {\bfseries John Smith} received his BSc (Mathematics) in 2000 from The University of Maryland. His research interests include lasers and optics.
\end{minipage}
\begin{minipage}{1.0\textwidth}
  \begin{wrapfigure}{L}{0.25\textwidth}
    \includegraphics[width=0.25\textwidth]{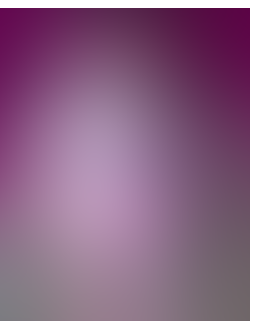}
  \end{wrapfigure}
  \noindent
  {\bfseries Alice Smith} also received her BSc (Mathematics) in 2000 from The University of Maryland. Her research interests also include lasers and optics.
\end{minipage}
\endgroup
}{}

\end{document}